# Low–Temperature Series for Renormalized Operators: the Ferromagnetic Square–Lattice Ising Model

**J. Salas** [*]
*Department of Physics*
*New York University*
*4 Washington Place*
*New York, NY 10003, USA*

**Abstract**

A method for computing low–temperature series for renormalized operators in the two–dimensional Ising model is proposed. These series are applied to the study of the properties of the truncated renormalized Hamiltonians when we start at very low temperature and zero field. The truncated Hamiltonians for majority rule, Kadanoff transformation and decimation for 2 × 2 blocks depend on the how we approach the first–order phase–transition line. These Renormalization Group transformations are multi–valued and discontinuous at this first–order transition line when restricted to some finite–dimensional interaction space.

[*]E-mail: salas@mafalda.physics.nyu.edu

# 1 Introduction

The behavior of the Renormalization Group (RG) in the vicinity of first–order phase transitions has been a very controversial matter for the last 20 years. In 1975 Nienhuis and Nauenberg [1] proposed that the RG transformations (RGTs) behave near first–order transition points in a similar fashion as near standard critical points. Each RG step is smooth (i.e. the renormalized couplings are analytic functions of the original ones, even at the transition points). Singular behavior is recovered as we infinitely iterate this transformation near the attracting manifold of a fixed point. Moreover, the fixed point governing a first–order phase transition has the following properties [1]: i) A domain of attraction which includes the first–order transition surface. ii) Zero correlation length at the fixed point (In most systems, first–order transition points possess a finite correlation length. See [2] for a counterexample). iii) A relevant operator whose critical exponent is given by the dimensionality of the system, $y = d$. As a matter of fact, there are as many exponents $y = d$ as phases coexist at the transition line [2] [3]. In the Ising model it is believed that this fixed point is located at zero temperature [4].

This picture was criticized by some authors [5,6,7,8,9] who claimed that the RG flow is itself discontinuous at the transition line. That is, they claimed that the renormalized Hamiltonian has different limiting values depending on how the original Hamiltonian approaches the transition line. Most of these claims were based on Monte Carlo Renormalization Group (MCRG) computations [5,6,7,8]. In ref. [9] non–rigorous analytical arguments were given to support the same conclusion.

In opposition to this view, it was argued in ref. [10] that the observed discontinuities are due to the truncation of the Hamiltonian space inherent in the MCRG approach. That is, the exact RG map is continuous at the first–order phase transition (in agreement with the conventional scenario), but truncation could induce the observed discontinuities. In fact, for the two–dimensional (2D) Ising model and majority rule with $2 \times 2$ blocks it was found that the discontinuity in the magnetic field was of the same order as the truncation error. Moreover, as the number of operators included in the computation was increased, the size of this discontinuity decreased.

This puzzle was partially solved by van Enter–Fernández–Sokal [11], who showed that for systems with bounded dynamical variables and interacting through a Hamiltonian belonging to the space $\mathcal{B}^1$ (i.e. the space of real, absolutely summable and translation–invariant interactions) the RG map is always continuous and single–valued, *whenever it exists at all* (subject to some very mild locality conditions on the RGT). For finite systems the existence of the transformation (i.e. of the renormalized Hamiltonian) is trivial. In the thermodynamic limit, however, this is a very subtle problem. As a matter of fact, van Enter *et al* proved that the renormalized Hamiltonian does *not* exist in the 2D Ising model when the temperature is low enough, for the Kadanoff transformation, decimation, block average and some particular cases of majority rule. The possibility of a non–Gibbsian renormalized measure had been explained earlier by Griffiths and

---

[1] Nienhuis and Nauenberg called this a discontinuity fixed point (DFP). However, we shall avoid this name as it causes confusion with a different scenario (discontinuous RG flow) described in the next paragraph.

[2] Here we take into account the (trivial) critical exponent $y = d$ associated with the renormalization of the identity operator in the Hamiltonian.

Pearce [12] using more physical (but non–rigorous) arguments. On the other hand, what happens for the majority rule with $2 \times 2$ blocks (the case most considered in the literature) is still unclear.

It therefore remains to be understood (a) why MCRG methods show a discontinuous RG flow at or near first–order phase transition points and (b) whether such apparent discontinuities are in any way connected with the Gibbsian or non–Gibbsian nature of the renormalized measure. The logic of the MCRG approach is the following: 1) Compute expectation values of suitable renormalized local operators using a Monte Carlo (MC) algorithm and a certain real–space RGT. 2) *Assume* that the renormalized measure is a Gibbs measure for some (a priori unknown) interaction $\mathcal{H}'_n$ [3] belonging to a (pre–chosen) $n$–dimensional subspace $V_n \subset \mathcal{B}^1$. 3) Compute the renormalized coupling constants (i.e. compute $\mathcal{H}'_n$) using some statistical inversion method. In general, one has to minimize some strictly convex functional $F_n$ [4], which leads to a highly non–linear set of equations involving the expectation values computed in step 1. MCRG methods thus study a truncated RG map in which the renormalized Hamiltonian $\mathcal{H}'_n$ is forced to lie in the subspace $V_n$. At this point we can ask two different questions: (A) For each fixed $n < \infty$, what are the properties of $\mathcal{H}'_n$: existence, uniqueness, continuity,...? (B) What is the behavior of $\mathcal{H}'_n$ as $n$ tends to infinity? Presumably, the answers to (A) and (B) are model–dependent: they depend generically on the physical system, on the RGT and on the choice of the subspaces $V_n$. On the other hand, the connection between Gibbsianness and the behavior of the $\mathcal{H}'_n$ as $n \to \infty$ is far from clear. In [11] it was proven that *if the renormalized measure is Gibbsian*, then there is a sequence of truncated interactions $\{\widehat{\mathcal{H}'_n}\}$, which are almost minimizers of $F_n$ and which converge to the true renormalized Hamiltonian $\mathcal{H}'$ in $\mathcal{B}^1$ norm. However, it is not guaranteed that the exact minimizers $\mathcal{H}'_n$ also converge to $\mathcal{H}'$. On the other hand, *if the renormalized measure is non–Gibbsian*, then it was proven in [11] that the the sequence $\{\mathcal{H}'_n\}$ has no limit at all in the space $\mathcal{B}^1$.

The simplest system which undergoes a first–order phase transition is the 2D Ising model at $T < T_c$. In particular, for $T \ll T_c$ we can do a complete analytical study using low–temperature (low–$T$) expansions [13] [5]. Analytical methods would be preferred in this kind of investigation, as all sources of errors are under control [6,7].

In this paper we address question (A) for the 2D Ising model at low temperature, using series expansions. We develop a simple procedure to compute these series for the expectation values of the renormalized operators which enter in the equations needed to obtain $\mathcal{H}'_n$. For real–space RGTs the expectation value of an operator $O$ with respect the renormalized measure can be written as an expectation value in the original measure of a certain composite operator $\tilde{O}$. This $\tilde{O}$ is equal to the original operator $O$ acted upon by a probability kernel (which is the mathematical object representing the RGT). Thus, if we know how to obtain the low–$T$ expansions in the original (unrenormalized) measure, then we can compute any expectation value by doing the corresponding integral.

---

[3]We shall hereafter denote renormalized quantities with a prime.

[4]In ref. [11], Section 5.1.2, the interaction $\mathcal{H}'_n$ comes from minimizing the relative entropy density with respect to the renormalized measure. See also Section 3 below.

[5]It ought to be also possible to study the first–order transition undergone by the $q$–state Potts model in the limit $q \gg 1$ using $1/q$ expansions.

[6]Series expansions do not suffer from two types of errors inherent to any MC simulation: statistical errors and finite–size corrections. These errors affect the estimators of the expectation values obtained in the MC simulation and they are propagated to the interaction $\mathcal{H}'_n$. By contrast, the predictions for such expectation values given by series expansions are obtained directly in the thermodynamic limit and no stochastic process is involved.

[7]Note that unlike many applications of series expansions, here we are really interested in the low–$T$ behavior and *not* in the critical region $T \approx T_c$. Therefore, no extrapolation procedure is involved.

These series can be useful in two other ways. They provide a check for MCRG computations at low temperature, since expectation values coming from the MC simulations can be compared with the low–$T$ predictions. On the other hand, when performing a RGT the system is viewed at a larger spatial scale. For that reason we believe that the low–$T$ series for the renormalized magnetization, susceptibility and specific heat could be used to extract the critical exponents (using standard series–extrapolation techniques). In fact, a better convergence could be expected for these "improved" series. It would be interesting to devise a computational procedure to generate these series to an arbitrary order.

The goal of this paper is thus to study the properties of the finite–dimensional approximants $\mathcal{H}'_n$ for the 2D Ising model. Starting on the first–order transition line (i.e. at zero magnetic field) and at very low temperature, for each of the two pure phases $\nu^{(\pm)}$ we obtain (via some RGT) two different renormalized measures $\nu'^{(\pm)}$. For each of them we can find a unique truncated renormalized Hamiltonian $\mathcal{H}'^{(\pm)}_n$. The natural question is: are these two truncated Hamiltonians equal or not? Or equivalently: do all the odd couplings in $\mathcal{H}'^{(\pm)}_n$ vanish, or not? An affirmative answer implies that the truncated RGT restricted to $V_n$ is continuous and single–valued at the transition line [8].

We have studied three different RGTs: decimation, Kadanoff transformation with large parameter $p$ and majority rule, all of them defined on $2 \times 2$ blocks [9]. We find that the truncated Hamiltonian $\mathcal{H}'_n$ is continuous at the transition line for the majority–rule transformation when restricted to a subspace containing a magnetic field and a nearest–neighbor interaction. On the other hand, we find that this is not the case for the decimation and large–$p$ Kadanoff transformations restricted to the latter 2D subspace nor for the majority–rule transformation when restricted to the three–dimensional subspace containing magnetic field, nearest–neighbor and next–to–nearest–neighbor interactions. In all these latter cases, the renormalized magnetic field is *non–zero*, implying that the truncated RG map is *discontinuous* at the first–order transition line. Thus, the typical situation seems to be that truncation induces discontinuities in the RG map when restricted to some finite–dimensional subspace of $\mathcal{B}^1$.

This paper is organized as follows. In Section 2 we explain how the low–$T$ expansions for renormalized observables can be obtained. We give three examples for the 2D Ising model: decimation, Kadanoff transformation and majority rule, all of them with block size $b = 2$. In Section 3 we study of these RGTs near the Ising first–order phase–transition. Finally in Section 4 we present our conclusions.

---

[8] In this case, the truncated Hamiltonian $\mathcal{H}'_n = \mathcal{H}'^{(\pm)}_n$ satisfies a fundamental property of the exact renormalized interaction $\mathcal{H}'$ (if $\mathcal{H}'$ exists at all), namely single–valuedness and continuity.

[9] It is known [11] that the first two transformations lead to non–Gibbsian renormalized measures at very low temperatures. This question is not clear for the last transformation, although it has been conjectured that it is also non–Gibbsian. Unfortunately, we are not aware of any real–space local transformation leading to a Gibbsian measure for the Ising model at low temperatures. In any case, we limit ourselves to studying question (A), for which the Gibbsianness or non–Gibbsianness is unlikely to play a significant role.

## 2 Series Expansions for Renormalized Operators

### 2.1 Review of Low-$T$ Expansions

Let us consider for simplicity a ferromagnetic Ising model on a 2D square lattice. The spins take the values $\pm 1$ and interact through the Hamiltonian

$$\mathcal{H} = -K \sum_{\langle i,j \rangle} (\sigma_i \sigma_j - 1) - H \sum_i (\sigma_i - 1) \tag{2.1}$$

where the first sum is over all the nearest–neighbor pairs of spins, and the second one over every point $i = (i_x, i_y)$ of the lattice. The partition function for a system of $N$ spins with periodic boundary conditions is then

$$Z_N = \sum_{\{\sigma = \pm 1\}} \exp\left[ K \sum_{\langle i,j \rangle} (\sigma_i \sigma_j - 1) + H \sum_i (\sigma_i - 1) \right] \tag{2.2}$$

We have absorbed the term $\beta = 1/kT$ in the definition of the coupling constants $K \geq 0$ and $H$. We are mainly interested in the zero–field case ($H = 0$), but for future convenience we keep the second term of the Hamiltonian (2.1). This term will be necessary to obtain the zero–field magnetization (see below).

The first step to compute low–$T$ expansions is to find out the ground states of the system at $T = 0$. In our case it is easy to realize that when $H = 0$ there are only two translation–invariant ground states. Both of them are completely ordered configurations with magnetization $+1$ and $-1$ respectively. When $H \neq 0$ then there is only one ground state whose magnetization is parallel to the magnetic field $H$. We will choose hereafter the $(+1)$–state as our ground state. This implies that the magnetic field should be always non–negative ($H \geq 0$). Furthermore, we have normalized the Hamiltonian (2.1) in such a way that $\mathcal{H}(+1) = 0$.

Looking at eq. (2.2) it is easy to realize that each flipped spin is penalized by a factor $\lambda = \exp(-2H)$ in the partition function. And each unsatisfied bond (i.e. a bond with both spins in opposite states) is suppressed by a factor $\mu = \exp(-2K)$. All the spin configurations with $n$ flipped spins and $m$ unsatisfied bonds give the same contribution to the partition function (2.2) and equal to $\mu^m \lambda^n$. So we can group these configurations together and express the partition function as

$$Z_N(\mu, \lambda) = \sum_{m,n} Z_{m,n}^{(N)} \mu^m \lambda^n \tag{2.3}$$

where $Z_{m,n}^{(N)}$ is the number of configurations with $m$ unsatisfied bonds and $n$ flipped spins that occur in the system. These numbers depend explicitly on the size of the system, as well as on the boundary conditions. The first term of the expansion corresponds to the ground state, the second to one flipped spin ($n = 1$, $m = 4$), the third to two nearest–neighbor flipped spins ($n = 2$, $m = 6$), and so on. With this choice of boundary conditions, $Z^{(N)}_{m,n} = 0$ for odd values of $m$. This expansion is exact for finite $N$ if all the $2^N$ possible configurations are taken into account.

The low-$T$ expansion of the partition function (2.3) contains the most relevant terms when the temperature goes to zero. It can also be viewed as an enumeration of the low–energy excitations of the system. Here we are interested in developing an expansion valid as $K \to \infty$ with $H$ bounded (i.e. an expansion in powers of $\mu$ ($\ll 1$) whose coefficients are functions of $\lambda$) [10]. Thus, the dominant terms are those with the smallest values of $m$. For a given value of $m$ the possible values of $n$ are finite. For excitations which do not see the boundary of the system the allowed values of $n$ are given by $n \in [m/4, m^2/16] \cup [N - m^2/16, N - m/4]$ (resp. $[(m+2)/4, (m^2-4)/16] \cup [N - (m^2-4)/16, N - (m+2)/4]$) when $m/2$ is even (resp. odd). All the terms with the same $m$, irrespective of $n$, are considered to contribute at the same order (i.e. $\lambda$ is considered to be of order 1). This feature implies that we can compute derivatives of the series expansions with respect to the magnetic field $H$. When the temperature is very close to zero only a few terms are needed to provide an accurate description of the system.

Actually, the partition–function expansion is a technical tool to compute the expectation values of some local operators: the energy density $E = \langle \sigma_{(0,0)} \sigma_{(1,0)} \rangle$ and the magnetization $M = \langle \sigma_{(0,0)} \rangle$. The relations for a finite system are the following

$$E_N(\mu, \lambda) = 1 + \frac{1}{2N} \frac{1}{Z_N} \frac{\partial Z_N}{\partial K} = 1 - \frac{1}{N} \frac{\mu}{Z_N} \frac{\partial Z_N}{\partial \mu} = \sum_{m,n} E^{(N)}_{m,n} \mu^m \lambda^n \qquad (2.4\text{a})$$

$$M_N(\mu, \lambda) = 1 + \frac{1}{N} \frac{1}{Z_N} \frac{\partial Z_N}{\partial H} = 1 - \frac{1}{N} \frac{2\lambda}{Z_N} \frac{\partial Z_N}{\partial \lambda} = \sum_{m,n} M^{(N)}_{m,n} \mu^m \lambda^n \qquad (2.4\text{b})$$

As before, the coefficients $\{E^{(N)}_{m,n}, M^{(N)}_{m,n}\}$ do depend on the lattice size and, in general, on the boundary conditions.

Let us discuss now the thermodynamic limit ($N \to \infty$) of these expansions. In this limit, the contribution of all the terms with the same $m$ is not in general of the same order. In particular, for $H > 0$ the configurations with $n$ near $N$ (for instance, $n \in [N - m^2/16, N - m/4]$ for $m/2$ even) are exponentially suppressed, and can therefore be dropped. Moreover, for $H = 0$ the $\sigma \to -\sigma$ symmetry implies that the contribution of the terms with $n$ near zero is equal to the one of those with $n$ near $N$. However, at $H = 0^+$ only the first set is selected. Therefore, for $H > 0$ or $H = 0^+$ the correct expansion is obtained by taking all the terms with $n$ near zero.

On the other hand, the series corresponding to the partition function (2.3) are meaningless when $N \to \infty$, as all the coefficients $Z^{(N)}_{m,n}$ (except for $Z^{(N)}_{0,0} = 1$) diverge in that limit. This is not true for the series (2.4a,2.4b) whose coefficients have a well-defined limit

$$E(\mu, \lambda) = \sum_{m,n} E_{m,n} \mu^m \lambda^n \quad ; \quad E_{m,n} = \lim_{N \to \infty} E^{(N)}_{m,n} \qquad (2.5\text{a})$$

$$M(\mu, \lambda) = \sum_{m,n} M_{m,n} \mu^m \lambda^n \quad ; \quad M_{m,n} = \lim_{N \to \infty} M^{(N)}_{m,n} \qquad (2.5\text{b})$$

---

[10] Different expansions are obtained when $H \to \infty$ and $K$ remains bounded or when both $K$ and $H$ diverge with $K/H \to$ constant.

The limiting coefficients $\{E_{m,n}, M_{m,n}\}$ do not depend on the boundary conditions of the finite systems. It is therefore reasonable to expect that $\{E_{m,n}, M_{m,n}\}$ are the coefficients of the true infinite–volume series [11].

The series expansions for the zero–field case ($H = 0^+$ or $\lambda = 1^-$) can be easily obtained from the previous ones by summing over the index $n$. For example, $M(\mu) = \sum_m M_m \mu^m$ where $M_m = \sum_n M_{m,n}$.

In this paper we are mainly concerned about the computation of expectation values of more complicated local observables $O$. By local operator we mean an operator which only depends on a finite number of spins. Our definitions of the energy density and the magnetization do satisfy this property. The previous procedure can be generalized to include also this case by adding to the Hamiltonian (2.1) a new term proportional to a translation–invariant version of the operator $O$.

However, this method is not feasible for very complicated operators, such as the ones considered in the next Section. In this paper we propose to use the definition

$$\langle O \rangle = \lim_{N \to \infty} \frac{1}{Z_N} \sum_{\{\sigma = \pm 1\}} O(\sigma)\, \mathrm{e}^{-\mathcal{H}(\sigma)} \qquad (2.6)$$

to overcome this problem. The term $\exp(-\mathcal{H})$ can be expanded in terms of configurations with $m$ unsatisfied bonds and $n$ flipped spins as we did in (2.3). In this case not all the configurations with the same values of $m$ and $n$ give the same contribution to the numerator of (2.6). This contribution is equal to $\mu^m \lambda^n$ times the value of the operator $O(\sigma)$ at the configuration. Let us consider a simple example. To compute the magnetization series one has to consider, for instance, the operator $O = \sigma_{(0,0)}$ (translation invariance assures that the mean value of this operator will coincide with the magnetization (2.5b)). For instance, the contribution of the one–flip configurations is different depending on whether the flipped spin coincides or not with $\sigma_{(0,0)}$. In the first case it is equal to $-\mu^4 \lambda$ and in the second one to $+\mu^4 \lambda$. The same occurs for more complicated configurations (and operators). For a finite volume we obtain in this way an expansion similar to (2.4a, 2.4b). The final result $\langle O \rangle = \sum_{m,n} O_{m,n} \mu^m \lambda^n$ is obtained after performing the thermodynamic limit.

The main advantage of this method is that it allows the computation of low–$T$ series for arbitrary operators. Its main drawback is that we need to compute two series for each observable, not one as in the former method.

## 2.2 Renormalization Group Transformations

Let us begin by considering RGT as a map from an (infinite–volume) Gibbs measure $\nu$ to a renormalized measure $\nu'$. Later on, the relationship between the measure $\nu'$ and the renormalized Hamiltonian $\mathcal{H}'$ will be discussed.

---

[11] A non–trivial interchange of limits is involved here, but it can presumably be justified rigorously by standard mathematical–physics techniques. We also need to be able to differentiate the free energy with respect to $\lambda$ at $\lambda = 1^-$, and this might be problematic since the free energy is not analytic in $H$ at $H = 0$ [14]. However, this function is infinitely differentiable, so we expect that our results are correct, as we compute everything in the stable phase. These issues arise in *all* uses of the conventional low–$T$–expansion technology, not only our own.

The first step is to define the renormalized spins. We divide the whole lattice into blocks (for simplicity we will assume here that these are 2 × 2 blocks). To each block $B_i$ we associate a new (renormalized) spin $\sigma'_i$. The RGT is the rule which gives the $\{\sigma'\}$ configuration from the original one $\{\sigma\}$. This rule could be either deterministic or stochastic, but in any case the renormalized spin should only depend on the spins belonging to the corresponding block (strict locality condition). Mathematically speaking we give a probability kernel $T(\sigma, d\sigma')$. For each configuration of the original spins $\{\sigma\}$, $T(\sigma, \cdot)$ is a probability distribution for the $\{\sigma'\}$ spins and furthermore, it satisfies the property $\int T(\sigma, d\sigma') = 1$. We assume that $T$ is strictly local in position space and that it maps translation–invariant measures into translation–invariant ones.

The probability distribution $\nu'$ of the image system is therefore given by

$$\nu' = \nu T = \int d\nu(\sigma) T(\sigma, \cdot) \tag{2.7}$$

and the expectation value of any local observable in this renormalized measure can be written as

$$\langle O(\sigma') \rangle_{\nu'} = \int d\nu(\sigma) \left[ \int T(\sigma, d\sigma') O(\sigma') \right] = \langle \tilde{O}(\sigma) \rangle_\nu \tag{2.8}$$

Thus, the probability kernel $T$ when acting (to the left) on the measure $d\nu(\sigma)$ yields a probability distribution on the new spins $\{\sigma'\}$ (i.e. a renormalized measure $\nu'$). On the other hand, when $T$ acts (to the right) on the operator $O(\sigma')$ the result is a composite operator $\tilde{O}(\sigma) = (T \cdot O)(\sigma)$ which depends only on the original spins. The expectation value of any local operator in the renormalized measure is equal to the mean value of a certain composite operator in the original measure.

This discussion is general: the conclusions hold whether or not the systems can be described by a Hamiltonian $\mathcal{H} \in \mathcal{B}^1$. Now we take into account the role of the Hamiltonians. Given an interaction $\mathcal{H} \in \mathcal{B}^1$ we can construct a measure on the spin configuration space using the Gibbs prescription

$$d\nu(\sigma) = d\nu^0(\sigma) \frac{1}{Z} e^{-\mathcal{H}(\sigma)}, \tag{2.9}$$

where $d\nu^0(\sigma)$ is the a–priori measure we assign to the space of configurations of a single spin (in our case it is just the counting measure which gives to each state a probability 1/2). For finite systems, formula (2.9) gives the correct answer; but for infinite systems one has to be more careful and consider the limit of the finite–volume measures with given boundary conditions as the system size tends to infinity. For finite systems the relation between Hamiltonians and measures is one–to–one. However, in the thermodynamic limit this is not the case: one Hamiltonian can be associated to several measures (i.e. at first order phase transitions), and conversely there are perfectly sound measures which cannot be constructed via the Gibbs prescription from any Hamiltonian $\mathcal{H} \in \mathcal{B}^1$ [11].

The Hamiltonian (2.1) does obviously belong to the set $\mathcal{B}^1$, so we can construct the measure $\nu$ using (2.9). Then the expectation value (2.8) of any local renormalized operator can be written as

$$\langle O \rangle_{\nu'} = \langle \tilde{O} \rangle_\nu = \lim_{N \to \infty} \frac{1}{Z_N} \sum_{\{\sigma = \pm 1\}} \tilde{O}(\sigma) e^{-\mathcal{H}(\sigma)} \tag{2.10}$$

Here any given choice of boundary conditions gives rise to an original Gibbs measure $\nu$ and a corresponding Gibbs measure $\nu'$.

In Section 2.1 we showed how to obtain low–$T$ expansions for a general mean value $\langle O \rangle_\nu$. Thus, the same procedure can be applied to (2.10), and series of the type $\langle O \rangle_{\nu'} = \sum_{m,n} O'_{m,n} \mu^m \lambda^n$ are obtained. The practical applicability of this method relies heavily on the actual form of the kernel $T$ as it is shown below. This procedure can also be easily generalized to several RG steps.

It is important to remark that this method does not suffer from any of the pathologies which are exhibited by the RG when we try to define it as a map from a Hamiltonian space into a Hamiltonian space. Here we have not tried to define any renormalized interaction $\mathcal{H}'$ related with the renormalized measure $\nu'$ via the Gibbs prescription (2.9). Our results are independent of the Gibbsian or non–Gibbsian nature of the renormalized measure. Let us illustrate this method with three examples:

**Example 1: Decimation**

This case is really simple because this transformation fixes one spin of the block to be the renormalized one. In particular, the (deterministic) kernel $T$ takes the form

$$T(\sigma, \sigma') = \prod_i \delta(\sigma'_i, \sigma_{2i}) \tag{2.11}$$

where the product is over all sites $i$ of the renormalized system.

We are only interested in computing observables that are monomials of the spins ($O = \{\sigma_{(0,0)}, \sigma_{(0,0)}\sigma_{(1,0)}\}$). So it is enough to compute for each RGT the composite operator $\tilde{\sigma}_i$. In this case this is equal to $\tilde{\sigma}_i = \int T(\sigma, d\sigma')\sigma'_i = \sigma_{2i}$. This implies that this case is trivial: the renormalized correlation functions are equal to the unrenormalized ones at twice the distance. These functions can be obtained in the 2D Ising model from the exact solution [15,16].

**Example 2: Kadanoff Transformation**

This is given by the following (stochastic) probability kernel

$$T(\sigma, \sigma') = \prod_i \frac{e^{p\sigma'_i \sum_{j \in B_i} \sigma_j}}{2\cosh(p \sum_{j \in B_i} \sigma_j)} \tag{2.12}$$

where $p$ is a free real parameter. Then, $\tilde{\sigma}_i = \tanh\left(p \sum_{k \in B_i} \sigma_k\right)$. The first terms can be computed by hand

$$\begin{aligned} M'(\mu, 1^-) &= \tanh 4p - 4(\tanh 4p - \tanh 2p)\mu^4 - 4(3\tanh 4p - 2\tanh 2p)\mu^6 \\ &\quad - (36\tanh 4p - 4\tanh 2p)\mu^8 + \mathcal{O}(\mu^{10}) \end{aligned} \tag{2.13a}$$

$$\begin{aligned} E'(\mu, 1) &= \tanh^2 4p - 8(\tanh^2 4p - \tanh 4p \tanh 2p)\mu^4 \\ &\quad - 2(11\tanh^2 4p - 6\tanh 2p \tanh 4p - \tanh^2 2p)\mu^6 \\ &\quad - (43\tanh^2 4p + 40\tanh 2p \tanh 4p - 20\tanh^2 2p)\mu^8 + \mathcal{O}(\mu^{10}) \end{aligned} \tag{2.13b}$$

The limit $p \to 0$ corresponds to the case in which the $\sigma'$ are not correlated with the original spins and thus, the renormalized spins do not interact among them. For this reason both quantities are zero. The limit $p \to \infty$ corresponds to the majority rule with equally–probable tie–breaker. This case will be treated in the next section.

**Example 3: Majority Rule**

In this case

$$T(\sigma', \sigma) = \prod_i \delta \left( \sigma'_i - \text{sign} \left( \sum_{j \in B_i} \sigma_j \right) \right) \tag{2.14}$$

When $\text{sign}(\cdot) = 0$ we choose $\sigma' = -1$ or $+1$ with probabilities $q \in [0,1]$ and $1 - q$ respectively. The composite operator $\tilde{\sigma}$ takes the form $\tilde{\sigma}_i = \text{sign}\left( \sum_{k \in B_i} \sigma_k \right)$. The first terms for general $q$ are:

$$M'(\mu, 1^-) = 1 - 8q\mu^6 - (10 + 44q)\mu^8 + \mathcal{O}(\mu^{10}) \tag{2.15a}$$
$$E'(\mu, 1) = 1 - 16q\mu^6 - (20 + 88q - 4q^2)\mu^8 + \mathcal{O}(\mu^{10}) \tag{2.15b}$$

The result with $q = 1/2$ was first reported in ref. [11]. Notice that the $\mathcal{O}(\mu^4)$ term vanishes. This is due to the fact that one–spin excitations cannot produce any flipped renormalized spin $\sigma' = -1$.

# 3 Study of the First–Order Phase Transition at Very Low Temperatures

In MCRG calculations one chooses in advance a linear subspace $V_n \subset \mathcal{B}^1$ of the space of local Hamiltonians. Then, given certain renormalized expectation values, one tries to find a renormalized Hamiltonian $\mathcal{H}'_n \in V_n$ in such a way that a measure constructed from $\mathcal{H}'_n$ is similar in some sense to the true renormalized measure $\nu'$. Most "reconstruction" methods are based on Schwinger–Dyson equations [17,18,19]. The idea is simple: minimize a certain functional (which depends on the method) involving both renormalized expectation values (the input) and renormalized couplings (the output). It can be shown [19] that these methods provide a unique solution $\mathcal{H}'_n$, which coincides with the true one $\mathcal{H}'$ if this latter interaction belongs to the trial subspace $V_n$. The key property of these functionals is that they are strictly convex.

Here we will consider the procedure given in ref. [11], Section 5.1.2. It is based on the minimization of the relative entropy density with respect to the true renormalized measure $\nu'$. This functional in also strictly convex and thus, the solution is unique in each $V_n$ if it exists. Van Enter *et al.* showed that the solution $\mathcal{H}'_n$ should satisfy the following conditions

$$\langle O_i \rangle_{\nu'} = \langle O_i \rangle_{\nu'_n} \qquad \forall O_i \in V_n \tag{3.1}$$

where $\nu'_n$ is some Gibbs measure constructed from the Hamiltonian $\mathcal{H}'_n$. In this case we have the same number of equations as the number of unknown parameters. Note that $\mathcal{H}'_n$ might have multiple Gibbs measures; it is required that *one* of them satisfy (3.1).

First–order phase transitions are characterized by the coexistence of several pure phases. Given one RGT, each pure phase $\nu^{(k)}$ is mapped to a renormalized measure $\nu'^{(k)}$. Given a subspace $V_n \in \mathcal{B}^1$, we can solve eq. (3.1) for each renormalized measure $\nu'^{(k)}$ and obtain a corresponding truncated renormalized Hamiltonian $\mathcal{H}'^{(k)}_n$. All these truncated Hamiltonians are uniquely defined and the important question is whether they are all equal or not.

In the 2D Ising case, there are only two pure phases $\nu^{(\pm)}$ coexisting at $T < T_c$ and $H = 0$, and they are related by the $\sigma \to -\sigma$ symmetry. For this reason, the even couplings in the truncated Hamiltonians $\mathcal{H}'^{(\pm)}_n$ are equal, and the odd couplings differ by a sign (in particular, the renormalized magnetic field in one phase is minus the renormalized magnetic field in the other one). Therefore, the truncated RG map is continuous and single-valued if and only if the renormalized Hamiltonians do not contain any odd interaction. Due to this symmetry, we only have to consider one phase (i.e. the (+1)-phase of Section 2 [12] ). To check if the truncated Hamiltonian $\mathcal{H}'_n$ has any odd term we simply solve (3.1) restricted to the even-coupling subspace of $V_n$. If such a solution exists, then $\mathcal{H}'^{(+)}_n = \mathcal{H}'^{(-)}_n$; if no such solution exists, then $\mathcal{H}'^{(+)}_n \neq \mathcal{H}'^{(-)}_n$.

## Case I. $V_2 = \{H, K\}$

In this case [13] our subspace consists precisely of the Hamiltonians (2.1). We will try to match both the energy density and the zero-field magnetization by using a zero-field Hamiltonian at a different (lower) temperature. If this matching can be performed, it would mean that the truncated RGT for the subspace $V_2$ is continuous at the transition line.

Let us consider first the majority-rule map with random tie-breaker. We define $K'$ as the nearest-neighbor coupling such that

$$E'(K, 0) = E(K', 0) \tag{3.2}$$

Using the result (2.15b) and the well-known expansion of the Onsager solution

$$E(\mu, 1) = 1 - 4\mu^4 - 12\mu^6 - 36\mu^8 + \mathcal{O}(\mu^{10}) \tag{3.3a}$$
$$M(\mu, 1^-) = 1 - 2\mu^4 - 8\mu^6 - 34\mu^8 + \mathcal{O}(\mu^{10}) \tag{3.3b}$$

we find that

$$\mu' = \sqrt{2}\mu^3 + \frac{63}{8\sqrt{2}}\mu^5 + \mathcal{O}(\mu^6) \tag{3.4}$$

The zero-field magnetization $M$ at this particular temperature is equal to

$$M(\mu', 1^-) = 1 - 4\mu^6 - \frac{63}{4}\mu^8 + \mathcal{O}(\mu^9) \tag{3.5}$$

and this expansion should be compared with the renormalized magnetization $M'(\mu, \lambda)$ given in (2.15a). We find that

$$M(\mu', 1^-) > M'(\mu, 1^-) \tag{3.6}$$

---

[12]To simplify the notation all the superscripts (+) will be dropped hereafter.

[13]One might consider [8] the even simpler case $V_1 = \{H\}$. But this case is trivial since any RGT satisfying $M(K, 0^+) \neq M'(K, 0^+)$ is necessarily discontinuous at $H = 0$ when restricted to the affine subspace $A_1 = V_1 + (0, K)$ with fixed $K \gg K_c$.

This equation means that we can account for the observed renormalized magnetization $M'(\mu, 1^-)$ with a system at zero field and $K' = -(1/2)\log \mu' \approx 3K - (1/4)\log 2$. This system is not in a pure phase, but in a mixed phase because the renormalized magnetization $M'(\mu, 1^-)$ lies strictly between $\pm M(\mu', 1^-)$. Thus, eq. (3.1) is satisfied by a measure $\nu_2'$ which is a convex combination of the two pure phases $\nu_{K'}^{(\pm)}$ characterizing the 2D Ising model at inverse temperature $K'$ and field $H' = 0^\pm$ [i.e. $\nu_2' = \alpha \nu_{K'}^{(+)} + (1-\alpha)\nu_{K'}^{(-)}$ for a suitable $\alpha \in (0,1)$].

The same game can be played with the other two RGTs considered in Section 2. The easiest case is the decimation transformation, where conclusions can be drawn for every $K > K_c$. In the 2D Ising model it is well–known that $\langle \sigma_{(0,0)}\sigma_{(1,0)} \rangle > \langle \sigma_{(0,0)}\sigma_{(2,0)} \rangle$ for $0 < K < \infty$. This implies immediately that $E'(K, 0) < E(K, 0)$, and hence that $K' < K$ if we take into account that $E(K, 0)$ is a strictly increasing function of $K$. On the other hand, the renormalized magnetization coincides with the unrenormalized one (i.e. the RG flow follows the lines of constant magnetization). And $M(K, 0^+)$ is also a strictly increasing function of $K$ for $K > K_c$. Combining both pieces we obtain that $M(K', 0^+) < M'(K, 0^+)$ for all $K > K_c$. This inequality is opposite to (3.6), because the direction of the RG flow is also opposite that of the majority–rule flow: it goes from low temperature to high temperature ($K' < K$). So, we have to increase the magnetic field to keep the magnetization constant, unless the magnetization at the starting point is zero. The latter condition only holds above the critical temperature. In summary, for any $K > K_c$ we cannot match the renormalized observables using a zero–field Hamiltonian.

For the Kadanoff transformation and large (but *finite*) $p$ the same result holds: one cannot match the energy densities and the magnetizations with a zero–field nearest–neighbor interaction. This can only be proved when $p$ is large enough. The reason is clear: the leading term of $E'$ is $\tanh^2 4p$ and if $p$ is not large, then the solution of (3.2) does not satisfy $\mu' \ll 1$ and the low–$T$ series for $\mu'$ are then meaningless.

For finite $p$ we can always choose $\mu_0$ such that for $\mu < \mu_0$ the leading term of $E'(\mu, 1)$ is dominated by a term which does not depend on $\mu$. Then

$$E'(\mu, 1) = 1 - 4e^{-8p} + \mathcal{O}(e^{-16p}) \tag{3.7}$$

if we choose $\mu_0 \sim \exp(-3p)$. The solution of eq. (3.2) is then

$$\mu' = e^{-2p} - \frac{3}{4}e^{-6p} + \mathcal{O}(e^{-10p}) \tag{3.8}$$

and

$$M(\mu', 1^-) = 1 - 2e^{-8p} - 3e^{-12p} + \mathcal{O}(e^{-16p}) \tag{3.9}$$

which should be compared with the expansion of the renormalized magnetization for $p$ very large and $\mu < \mu_0$

$$M'(\mu, 1^-) = 1 - 2e^{-8p} + \mathcal{O}(e^{-16p}) \tag{3.10}$$

We find that the leading term of both quantities is the same, but the next–to–leading term is different. In particular we find that $M'(\mu, 1^-) > M(\mu', 1^-)$, so we cannot match both $E'$ and $M'$ with a zero–field Ising interaction. This discussion is valid as long as $p$ is large but *finite*. When $p$ diverges the leading term of $1 - E'(\mu, 1)$ is proportional to $\mu^6$ and we re–obtain the result for the majority rule transformation with $q = 1/2$.

**Case II.** $V_3 = \{H, K, L\}$

Now we are considering a Hamiltonian with an additional diagonal–next–to–nearest neighbor term $L \sum \sigma_i \sigma_k$. First of all we have to compute the renormalized mean value of the next–to–nearest neighbor correlation function. The result for the majority rule with random tie–breaker is

$$F'(\mu, 1) = \langle \tilde{\sigma}_{(0,0)} \tilde{\sigma}_{(1,1)} \rangle = 1 - 4\mu^6 - 64\mu^8 - 336\mu^{10} - 1578\mu^{12} + \mathcal{O}(\mu^{14}) \qquad (3.11)$$

We also need the functions $H'$ and $E'$ to the same order in $\mu$. The result is [14]:

$$\begin{aligned} M'(\mu, 1^-) &= 1 - 4\mu^6 - 32\mu^8 - 168\mu^{10} - 816\mu^{12} + \mathcal{O}(\mu^{14}) & (3.12\text{a}) \\ E'(\mu, 1) &= 1 - 8\mu^6 - 63\mu^8 - 312\mu^{10} - 1328\mu^{12} + \mathcal{O}(\mu^{14}) & (3.12\text{b}) \end{aligned}$$

The second step is to write down the expressions for $\langle O_i \rangle_{\nu_n}$, $\forall O_i \in V_3$. The result for zero magnetic field is

$$\begin{aligned} E(\mu, \gamma, 1) &= 1 - 4\mu^4 \gamma^4 - 12\mu^6 \gamma^8 - 24\mu^8 \lambda^{12} - 32\mu^8 \lambda^{10} + 36\mu^8 \lambda^8 \\ &\quad - 40\mu^{10} \lambda^{16} + \mathcal{O}(\mu^8 \lambda^6) & (3.13\text{a}) \\ F(\mu, \gamma, 1) &= 1 - 4\mu^4 \gamma^4 - 16\mu^6 \gamma^8 - 36\mu^8 \lambda^{12} - 40\mu^8 \lambda^{10} + 36\mu^8 \lambda^8 \\ &\quad - 64\mu^{10} \lambda^{16} + \mathcal{O}(\mu^8 \lambda^6) & (3.13\text{b}) \\ M(\mu, \gamma, 1^-) &= 1 - 2\mu^4 \gamma^4 - 8\mu^6 \gamma^8 - 20\mu^8 \lambda^{12} - 24\mu^8 \lambda^{10} + 18\mu^8 \lambda^8 \\ &\quad - 40\mu^{10} \lambda^{16} + \mathcal{O}(\mu^8 \lambda^6) & (3.13\text{c}) \end{aligned}$$

where $\gamma = \exp(-2L)$. Now we have to find out a pair $(\mu', \gamma')$ such that

$$E(\mu', \gamma', 1) = E'(\mu, 1); \qquad F(\mu', \gamma', 1) = F'(\mu, 1) \qquad (3.14)$$

The solution to leading order is $\mu' = 4\mu^2 + \mathcal{O}(\mu^4)$ and $\gamma' = 1/(32\mu')^{1/4}(1 + \mathcal{O}(\mu'))$. This implies that $K' \approx 2K - \log 2 > 0$ and $L' \approx (5/8)\log 2 - K'/4 \approx (7/8)\log 2 - K/2 < 0$. So, as $K \to \infty$, $K'$ and $-L'$ also diverge. The latter relation between $\mu'$ and $\gamma'$ should be taken into account when computing the actual order of a given term in the expansion of the partition function $Z_N(\mu', \gamma', 1^-)$ and its derivatives. In our case, this implies that the first two excitations to the ground state are of order $\mu'^3$ and $\mu'^4$ respectively. We have considered here all the excitations up to order $\mathcal{O}(\mu'^6)$.

A straightforward computation leads to the next–to–leading terms

$$\begin{aligned} \mu' &= 4\mu^2 \left[ 1 - \frac{69}{16}\mu^2 + \sqrt{2}\mu^3 + \frac{17027}{512}\mu^4 + \mathcal{O}(\mu^5) \right] & (3.15\text{a}) \\ \gamma' &= \left( \frac{1}{32\mu'} \right)^{1/4} \left[ 1 + \frac{327}{256}\mu' - \frac{3}{16\sqrt{2}}\mu'^{3/2} + \frac{144177}{131072}\mu'^2 + \mathcal{O}(\mu'^{5/2}) \right] & (3.15\text{b}) \end{aligned}$$

The magnetization (3.13c) computed at the latter solution is equal to

$$M(\mu', \gamma', 1^-) = 1 - 4\mu^4 - 32\mu^8 - \frac{2689}{16}\mu^{10} + \mathcal{O}(\mu^{11}) < M'(\mu, 1^-) \qquad (3.16)$$

This implies that we cannot match the renormalized expectation values with a zero–field interaction belonging to $V_3$.

---

[14] We have computed this expansions by using a computer algorithm based on the Recursive Counting Method of ref. [20]. Further details can be obtained from the author.

# 4  Conclusions

In this note we have shown how to compute low temperature expansions for the expectation values of local operators computed in the renormalized measure. In particular we have analyzed three RGTs: decimation, Kadanoff transformation with large but finite parameter $p$, and majority rule with random tie–breaker. All of them are defined on $2 \times 2$ blocks. We have been able to compute the first terms of the series corresponding to the renormalized magnetization and nearest–neighbor two–point correlation function for all these transformations.

The main goal of this note was the analysis of the truncation issue in the Ising model. The unrenormalized system is located at the Ising first–order transition line at very low temperature ($H = 0, K \gg K_c$). For the three transformations considered we have found that we need a magnetic field to solve the matching equations (3.1) when we restrict the truncated Hamiltonian to belong to a certain finite–dimensional subspace of $\mathcal{B}^1$. In particular, for the decimation and Kadanoff transformations this matching cannot be performed when restricting the equations to $V_2$. For majority rule, in this case the equations admit a zero–field solution but when we consider the (larger) subspace $V_3$ we also need a magnetic field.

So its seems that truncation in the renormalized Hamiltonian induces some spurious odd operators (we have only found non–zero magnetic fields, but there is no reason why more complicated odd operators should not appear for larger subspaces $V_n$). So, these RGTs are discontinuous at the Ising transition line when restricted to some finite–dimensional subspace of the interaction space $\mathcal{B}^1$.

However, these results do not clarify the interplay between truncation and non–Gibbsianness. It is known [11] that the decimation and Kadanoff transformations lead to non–Gibbsian renormalized measures when we start at low enough temperature; and in these cases we have shown that the truncated RGT are discontinuous. For the majority rule the situation is less clear, as the nature of the renormalized measure is not known. The authors of ref. [11] conjectured that in this case the renormalized measure is also non–Gibbsian, but they were able to prove it only for certain special block sizes ($7 \times 7$, $41 \times 41$, ...). In any case, this model leads to a continuous truncated RGT for the subspace $V_2$, but a discontinuous one for $V_3$. It is an open question what happens for larger subspaces $V_n$.

It would be very interesting to find a transformation which leads to a Gibbsian measure at low temperatures. In this case we could isolate the effect of truncation from non–Gibbsianness. A systematic study of the behavior of the estimates $\mathcal{H}'_n$ is also an interesting problem, which deserves more attention in the future.

## Acknowledgements

We would like to thank A. Sokal for his encouragement and for illuminating discussions. We also acknowledge helpful comments by J.L.F. Barbón, R. Fernández and M. García Pérez. This research has been supported by a MEC(Spain)/Fulbright grant.

# References

[1] B. Nienhuis and M. Nauenberg, Phys. Rev. Lett. **35** (1975) 477.

[2] M. Asorey and J.G. Esteve, J. Stat. Phys. **65** (1991) 483.

[3] M.E. Fisher and A.N. Berker, Phys. Rev. **B26** (1982) 2507.

[4] W. Klein, D.J. Wallace and R.K.P. Zia, Phys. Rev. Lett. **37** (1976) 639.

[5] H.J. Blöte and R.H. Swendsen, Phys. Rev. Lett. **43** (1979) 799.

[6] C.B. Lang, Nucl. Phys. **B280 [FS18]** (1987) 255.

[7] A. González–Arroyo, M. Okawa and Y. Shimizu, Phys. Rev. Lett. **60** (1999) 487.

[8] K. Decker, A. Hasenfratz and P. Hasenfratz, Nucl. Phys. **B295 [FS21]** (1988) 21.

[9] A. Hasenfratz and P. Hasenfratz, Nucl. Phys. **B295 [FS21]** (1988) 1.

[10] A. González–Arroyo and J. Salas, Phys. Lett. **B261** (1991) 415.

[11] A.C.D. van Enter, R. Fernández and A.D. Sokal, J. Stat. Phys. **72** (1993) 879.

[12] R.B. Griffiths and P.A. Pearce, J. Stat. Phys. **20** (1979) 499.

[13] C. Domb, in *Phase Transitions and Critical Phenomena*, Vol. 3. Edited by C. Domb and M.S. Green (Academic Press, New York).

[14] S.N. Isakov, Comm. Math. Phys. **95** (1984) 427.

[15] B. Kaufman and L. Onsager, Phys. Rev. **76** (1949) 1244.

[16] C.N. Yang, Phys. Rev. **85** (1952) 808.

[17] R.H. Swendsen, Phys. Rev. Lett. **52** (1984) 1165.

[18] A. González–Arroyo and M. Okawa, Phys. Rev. Lett. **58** (1988) 2165.

[19] A. González–Arroyo and J. Salas, Phys. Lett. **B214** (1988) 418.

[20] M. Creutz, Phys. Rev. **B43** (1991) 10659.